\documentclass[onecolumn,showpacs,floatfix,12pt,nofootinbib]{revtex4}
\usepackage[dvips]{graphics,color}
\usepackage{epsfig}\usepackage{float}
\RequirePackage{amssymb}
\usepackage{makeidx}
\usepackage[brazil]{babel}
\usepackage[latin1]{inputenc}
\usepackage{graphicx}
\usepackage{indentfirst}
\usepackage{color}
\newcommand{\be}{\begin{equation}}
\newcommand{\ee}{\end{equation}}

\newcommand{\ba}{\begin{eqnarray}}
\newcommand{\ea}{\end{eqnarray}}

\begin{document}

\title{Brane-Wor(l)ds within Brans-Dicke Gravity\footnote{Essay wrote to the volume:
{\it The Problems of Modern Cosmology},
by Tomsk State Pedagogical University, on occasion of 50th
birthday of Prof. S.D. Odintsov.}}

\author{M. C. B. Abdalla$^{1}$}
\email{mabdalla@ift.unesp.br}
\author{M. E. X. Guimar\~aes$^{2}$}
\email{emilia@if.uff.br}
\author{J. M. Hoff da Silva$^{1}$}
\email{hoff@ift.unesp.br}

\affiliation{1. Instituto de F\'{\i}sica Te\'orica, Universidade
Estadual Paulista, Rua Pamplona 145 01405-900 S\~ao Paulo, SP,
Brazil}

\affiliation{2. Instituto de F\'{\i}sica, Universidade Federal
Fluminense, Niter\'oi-RJ, Brazil}

\pacs{04.50.+h; 98.80Cq}

\begin{abstract}
We review some recent results obtained from the application of the
Gauss-Codazzi formalism to brane-worlds models in the Brans-Dicke
gravity. The cases of 4-branes embedded in a six-dimensional with
and without $\mathbb{Z}_{2}$ symmetry are both analyzed.

\end{abstract}
\maketitle
\section{Introduction}

Soon after the establishment of the modern concepts of
brane-worlds models \cite{RS} it was realized their importance in
the study of gravitational systems \cite{MAART}, as well as their
application in the analysis of cosmological problems
\cite{GALERA}. For codimension one brane-world models there is a
very powerful tool --- the so-called Gauss-Codazzi formalism ---
developed in order to project the gravitational field equations
from the bulk to the brane \cite{JP}. This procedure allows the
exploration of brane cosmological signatures in a deep way.
Roughly speaking, the existence of extra dimensions, within
gravitational context, added new source terms to the brane
projected gravitational field equation. Obviously, this also
happens in the Brans-Dicke \cite{BD} gravity. Besides, new
``source terms'' appear due to the dynamics of the additional
gravitational  scalar field. These terms appearing in the
projected gravitational equations in the Brans-Dicke theory
certainly lead to new cosmological signatures. The reason to
analyze brane-world models in such scalar-tensorial theory rests on
the fact that, at least sufficiently high energies, the General
Relativity is not able to  fully describe mostly of the puzzled
gravitational phenomena \cite{GSW}. Apart of that, there is a
strong interplay between Brans-Dicke theory and the gravity theory
recovered from string theory at low energies \cite{STRI}. In this
vein, it is possible to obtain information about some systems in
string theory by the analysis in Brans-Dicke framework. The
relationship between the two theories is given by a
--- model depended --- relabel of the Brans-Dicke parameter.

From the topological point of view, the brane(s) in the standard
Randall-Sundrum model \cite{RS} is (are) performed by domain
wall(s) and the extra dimension is a $S^{1}/\mathbb{Z}_{2}$
orbifold. It is familiar for cosmologists, however, that domain
walls (with symmetry breaking scales greater than 1 MeV) are
problematic \cite{ZELDO} and should not appear in a complete
scenario. The program of an exotic compactification of extra
dimensions using topological defects continued with global cosmic
strings in General Relativity \cite{RUTH} and with local and
global cosmic strings in the Brans-Dicke gravity \cite{AG}.
Because of the defect used to generate the bulk-brane structure
(the cosmic string), these models hold in six-dimensions. Models
studied in the Brans-Dicke framework \cite{AG} present only one
transverse extra dimension (codimension one models), while the
brane has five-dimensions with topology given by
$\mathbb{R}^{4}\times S^{1}$. The presence of a transverse and a
curled dimension in the same model is called hybrid
compactification \cite{INDY}.

The aim of this work is to review the application of the
Gauss-Codazzi formalism for hybrid compactification models in
the Brans-Dicke framework. In Section II we develop the main
lines of this approach in the case where the spacetime is endowed
with $\mathbb{Z}_{2}$ orbifold symmetry. In the Section III, we
work in the case without such symmetry and in Section IV, we
conclude with some final remarks and possible applications. We
stress that the main results delivered here, as well as the
details, are somehow described in \cite{AG2,AG3}.

\section{Projected gravitational field equations in $\mathbb{Z}_{2}$ symmetric
brane-worlds}

From now on we consider the brane as a five-dimensional
submanifold embedded in a manifold of six-dimensions --- the bulk.
As remarked before, the motivations to work in such
dimensionality can be found in \cite{RUTH,AG} (and
references therein). Denoting the covariant derivative in the bulk
by $\nabla_{\mu}$ and the one associated to the brane by $D_{\mu}$
the Gauss equation reads \cite{WALD}
\be{^{(5)}\!R^{\alpha}_{\beta\gamma\delta}=^{(6)}\!\!\!R^{\mu}_{\nu\rho\sigma}q_{\mu}^{\;\alpha}q_{\beta}^{\;\nu}q_{\gamma}^{\;\rho}q_{\delta}^{\;\sigma}+
K^{\alpha}_{\;\gamma}K_{\beta\delta}-K^{\alpha}_{\;\delta}K_{\beta\gamma}},\label{1}\ee
where $q_{\mu\nu}=g_{\mu\nu}-n_{\mu}n_{\nu}$ is the induced metric
on the brane ($n_{\mu}$ being the orthogonal unitary vector along
to the extra transverse dimension) and
$K_{\mu\nu}=q_{\mu}^{\;\alpha}q_{\nu}^{\;\beta}\nabla_{\alpha}n_{\beta}$
is the extrinsic curvature, which gives information about the
embedding of the brane. Starting from equation (\ref{1}), it is
easy to see that the Einstein tensor on the brane in given by
\newpage \ba
^{(5)}\!G_{\beta\delta}&=&\left.^{(6)}\!G_{\nu\sigma}q_{\beta}^{\;\sigma}q_{\delta}^{\;\sigma}+^{(6)}\!\!\!R_{\nu\sigma}n^{\nu}n^{\sigma}q_{\beta\delta}+KK_{\beta\delta}
-K_{\delta}^{\;\gamma}K_{\beta\gamma}\right.\nonumber\\&-&\left.
\frac{1}{2}q_{\beta\delta}(K^{2}-K^{\alpha\gamma}K_{\alpha\gamma})-\tilde{E}_{\beta\delta},\right.
\label{2}\ea where
$\tilde{E}_{\beta\delta}=^{(6)}\!\!R^{\mu}_{\;\nu\rho\sigma}n_{\mu}n^{\rho}q_{\beta}^{\;\nu}q_{\delta}^{\;\sigma}$.
Now, taking into account that the relation between the Riemann,
Ricci and Weyl tensors in an arbitrary dimension $(n)$ is \ba
^{(n)}\!R_{\alpha\beta\mu\nu}&=&\left.
^{(n)}\!C_{\alpha\beta\mu\nu}+\frac{2}{n-2}\Big(\;
 ^{(n)}\!R_{\alpha[\mu}g_{\nu]\beta}-^{(n)}\!\!\!R_{\beta[\mu}g_{\nu]\alpha}
\Big)\right.\nonumber\\&-&\left.\frac{2}{(n-1)(n-2)}\;
^{(n)}\!Rg_{\alpha[\mu}g_{\nu]\beta},\right. \label{3}\ea we can
rewrite the equation (\ref{2}) in the form \ba
^{(5)}\!G_{\beta\delta}&=&\left.\frac{1}{2}\;
^{(6)}\!G_{\nu\sigma}q_{\beta}^{\;\nu}q_{\delta}^{\;\sigma}-
\frac{1}{10}\;^{(6)}\!Rq_{\beta\delta}-\frac{1}{2}\;^{(6)}\!R_{\nu\sigma}q^{\nu\sigma}q_{\beta\delta}+KK_{\beta\delta}
-K^{\gamma}_{\;\delta}K_{\beta\gamma}\right.\nonumber\\&-&\left.\frac{1}{2}q_{\beta\delta}(K^{2}-K^{\alpha\gamma}K_{\alpha\gamma})
-E_{\beta\delta},\right.\label{prep}\ea where
$E_{\beta\delta}=^{(6)}\!C^{\mu}_{\;\nu\rho\sigma}n_{\mu}n^{\rho}q_{\beta}^{\;\nu}q_{\rho}^{\;\sigma}$.

The idea, hereafter, is to express the geometric quantities of the
brane in terms of the stress-tensor and the scalar dynamics of the
bulk, in order to apply the Gauss-Codazzi formalism to the case in
question. With this purpose, we remember that the
Einstein-Brans-Dicke equation is given by \ba
^{(6)}\!G_{\mu\nu}&=&\left.\frac{8\pi}{\phi}T_{M\mu\nu}+\frac{w}{\phi^{2}}\Big(\nabla_{\mu}\phi\nabla_{\nu}\phi-
\frac{1}{2}g_{\mu\nu}\nabla_{\alpha}\phi\nabla^{\alpha}\phi\Big)\right.\nonumber
\\&+&\left.\frac{1}{\phi}\Big(\nabla_{\mu}\nabla_{\nu}\phi-g_{\mu\nu}\Box^{2}\phi\Big),\right.\label{4}\ea
where $\phi$ is the Brans-Dicke scalar field --- the dilaton
---, $w$ a dimensionless parameter and $T_{M\mu\nu}$ the matter
energy-momentum tensor, everything except $\phi$ and gravity, in
the bulk. The scalar equation of Brans-Dicke theory is given by
\ba \Box^{2}\phi=\frac{8\pi}{3+2w}T_{M}.\label{scalar} \ea
Inserting the equation (\ref{scalar}) into (\ref{4}) and founding
the Ricci tensor and the scalar of curvature, it is possible to
express the equation (\ref{prep}) as \ba
^{(5)}\!G_{\beta\delta}&=&\left.\frac{1}{2}\Bigg[
\frac{8\pi}{\phi}T_{M\nu\sigma}+\frac{1}{\phi}\nabla_{\nu}\nabla_{\sigma}\phi+\frac{w}{\phi^{2}}\nabla_{\nu}\phi\nabla_{\sigma}\phi
\Bigg](q_{\beta}^{\;\nu}q_{\delta}^{\;\sigma}-q^{\nu\sigma}q_{\beta\delta})\right.\nonumber\\&+&\left.\frac{2\pi}{5\phi}q_{\beta\delta}
T_{M}\Bigg(\frac{13+27w}{3+2w}\Bigg)-\frac{7w}{20\phi^2}q_{\beta\delta}\nabla_{\alpha}\phi\nabla^{\alpha}\phi+KK_{\beta\delta}-K^{\gamma}_{\;\delta}
K_{\beta\gamma}\right.\nonumber\\&-&\left.\frac{1}{2}q_{\beta\delta}(K^{2}-K^{\alpha\gamma}K_{\alpha\gamma})-E_{\beta\delta}.\right.\label{NG}\ea

In order to extract information about this system we have to
compute the quantities on the brane. It can be implemented by
taking the limit of the extra transverse dimension tending to the
brane, but we have to specify the behavior of the extrinsic
curvature under such limit. This is a central piece in the
application of the Gauss-Codazzi formalism and strongly depends
whether or not the spacetime is endowed with a $\mathbb{Z}_{2}$
symmetry. In the case where there is such symmetry, it is possible
to show, by application of distributional calculus tools, that the
extrinsic curvature in one side of the brane, $K_{\mu\nu}^{+}$, is
related with it's other side partner, $K_{\mu\nu}^{-}$ by (see
reference \cite{AG2} for all the details)
\be{K_{\mu\nu}^{+}=-K_{\mu\nu}^{-}=\frac{4\pi}{\phi}\Bigg(-T_{\mu\nu}+\frac{q_{\mu\nu}(1+w)T}{2(3+2w)}
\Bigg)},\label{5}\ee and
\be{K^{+}=K^{-}=\frac{2\pi}{\phi}\Bigg(\frac{w-1}{3+2w}\Bigg)T}.\label{6}\ee
The relation (\ref{5}) is the generalization of the so-called
Israel-Darmois matching conditions \cite{ID} to the Brans-Dicke
gravity. It is possible to split the matter stress-tensor in
\be{T_{M\mu\nu}=-\Lambda
g_{\mu\nu}+\delta(y)T_{\mu\nu}},\label{7}\ee and
\be{T_{\mu\nu}=-\lambda q_{\mu\nu}+\tau_{\mu\nu}},\label{8}\ee
where $\Lambda$ is the cosmological constant of the bulk and
$\lambda$ the tension of the brane\footnote{We remark that the
delta term appearing in such decomposition can lead to
complications in a complete cosmological scenario.}. Now,
substituting the equations (\ref{5})-(\ref{8}) into (\ref{NG}) we
obtain
 \ba
^{(5)}\!G_{\beta\delta}&=&\left.\frac{1}{2}\Bigg[\frac{1}{\phi}\nabla_{\nu}\nabla_{\sigma}\phi+\frac{w}{\phi^{2}}\nabla_{\nu}\phi\nabla_{\sigma}\phi
\Bigg](q_{\beta}^{\;\nu}q_{\delta}^{\;\sigma}-q^{\nu\sigma}q_{\beta\delta})+8\pi
\Omega\tau_{\beta\delta}-\Lambda_{5}q_{\beta\delta}\right.\nonumber\\&+&\left.8\Big(\frac{\pi}{\phi}\Big)^{2}\Sigma_{\beta\delta}-E_{\beta\delta}
\right. \label{9},\ea where
\be{\Omega=\frac{3\pi(w-1)\lambda}{\phi^{2}(3+2w)}}\label{10},\ee
\be{\Lambda_{5}=\frac{-4\pi\Lambda(21-41w)}{5\phi(3+2w)}+\Big(\frac{\pi}{\phi}\Big)^{2}\Bigg[\frac{7w}{20\pi^{2}}\nabla_{\alpha}\phi\nabla^{\alpha}\phi
+\frac{24(w-1)\lambda}{(3+2w)^{2}}[(w-1)\lambda+\tau]\Bigg]}\label{11}\ee
and
\be{\Sigma_{\beta\delta}=q_{\beta\delta}\tau^{\alpha\gamma}\tau_{\alpha\gamma}-2\tau^{\gamma}_{\;\delta}\tau_{\gamma\beta}+\Big(\frac{3+w}{3+2w}
\Big)\tau\tau_{\beta\delta}-
\frac{(w^{2}+3w+3)}{(3+2w)^{2}}q_{\beta\delta}\tau^{2}}.\label{12}\ee

The main property to be noted from equation (\ref{9}) is that we
do not recover Brans-Dicke gravity on the brane if the dilaton
depends only on the extra transverse dimension. Instead, Einstein
equation is recovered with subtle but important modifications
coming from both extra dimensions and dilaton dynamics. Hence,
equation (\ref{9}) can be used to extract deviations from usual
General Relativity. We refer again the reader to reference
\cite{AG2} for more analysis and comments on the implications of
the result encoded in (\ref{9})-(\ref{12}).

\section{Lifting the $\mathbb{Z}_{2}$ symmetry}

The $\mathbb{Z}_{2}$ symmetry has a multiple role in brane-worlds
scenarios \cite{AG3}. In what concerns to the gravitational
aspects it determines univocally the jump of the extrinsic
curvature across the brane (\ref{5}). In this vein, it is not
surprising that the absence of such symmetry makes the
calculations a little more involved. In this Section we shall
present the guidelines of how to project the Einstein-Brans-Dicke
equation on the brane without the $\mathbb{Z}_{2}$ symmetry. More
details can be found in reference \cite{AG3} for the Brans-Dicke
case and in \cite{JP2} for the context of General
Relativity\footnote{Actually, the reference \cite{AG3} is
a first generalization of the work presented in \cite{JP2} to the
Brans-Dicke framework.}. Let us start defining two quite important
tools which determine the mean value of any tensorial quantity,
say $X$, \be \langle X \rangle =\frac{1}{2}(X^{+}+X^{-}),
\label{13} \ee and the jump across the brane \be
[X]=X^{+}-X^{-},\label{14} \ee where $X^{\pm}$ are both limits of
$X$ approaching the brane from both $\pm$ sides. It is not
difficult to see that the quantities defined by equations
(\ref{13}) and (\ref{14}) lead to the algebra \ba [AB]=\langle A
\rangle [B]+[A]\langle B \rangle, \label{15}\\ \langle AB
\rangle=\langle A \rangle \langle B
\rangle+\frac{1}{4}[A][B]\label{16}. \ea

Note that from the Gauss equation (\ref{1}), we can write down
the Ricci tensor on the brane in a more convenient way \be
^{(5)}\!R_{\mu\nu}=Y_{\mu\nu}+KK_{\mu\nu}-K^{\lambda}_{\mu}K_{\lambda\nu}\label{17},\ee
where \be Y_{\mu\nu}\equiv
\frac{3}{4}\,^{(6)}\!\!R_{\alpha\beta}q_{\mu}^{\alpha}q_{\nu}^{\beta}+\frac{1}{4}\,^{(6)}
\!\!R_{\alpha\beta}q^{\alpha\beta}q_{\mu\nu}- \frac{1}{5}\,^{(6)}
\!\!Rq_{\mu\nu}+E_{\mu\nu}.\label{18} \ee In order to obtain the
projected equation on the brane, one needs to apply the limits
defined in (\ref{13}) and (\ref{14}) into (\ref{18}). Starting
with $[^{(5)}\!R_{\mu\nu}]$, one has \be
[^{(5)}\!R_{\mu\nu}]=0=[Y_{\mu\nu}]+\langle K \rangle
[K_{\mu\nu}]+[K]\langle K_{\mu\nu} \rangle-\langle
K_{[\mu}^{\;\;\alpha}\rangle[K_{\nu]\alpha}] \label{19}.\ee Now,
by using the same decomposition showed in equations (\ref{7}) and
(\ref{8}), the equations (\ref{5}) and (\ref{6}) give,
respectively
\be[K_{\mu\nu}]=-\frac{8\pi}{\phi}\Bigg(\tau_{\mu\nu}+\frac{q_{\mu\nu}}{2(3+2w)}((w-1)\lambda-(w+1)\tau)
\Bigg)\label{20},\ee and \be [K]=\frac{8\pi(w-1)}{2\phi
(3+2w)}(\tau-5\lambda)\label{21}.\ee Hence, in the light of
(\ref{20}) and (\ref{21}), the equation (\ref{19}) results in
\begin{eqnarray}
-\Big(\frac{8\pi}{\phi}\Big)^{-1}[Y_{\mu\nu}]&=&\left. \langle
K_{\alpha[\mu}\rangle\tau_{\nu]}^{\alpha}-\Bigg(\tau_{\mu\nu}+\frac{q_{\mu\nu}}{2(3+2w)}((w-1)\lambda-(w+1)\tau)
\Bigg)\langle K \rangle \right. \nonumber \\&+&\left.
\frac{(3(1-w)\lambda-(w+3)\tau)}{2(3+2w)}\langle
K_{\mu\nu}\rangle\right. .\label{22}
\end{eqnarray}

This last equation will be useful helping to find the mean value
of the extrinsic curvature. Firstly, however, let us derive the
full projected equation. The mean operator acting on (\ref{18})
gives \be \langle \,^{(5)}\!R_{\mu\nu}
\rangle=\,^{(5)}\!R_{\mu\nu}=\langle
Y_{\mu\nu}\rangle+\frac{1}{4}\Big([K][K_{\mu\nu}]-[K_{\mu}^{\;\;\alpha}][K_{\nu\alpha}]
\Big)+\langle K \rangle \langle K_{\mu\nu} \rangle-\langle
K_{\mu}^{\;\;\alpha} \rangle \langle K_{\nu\alpha}
\rangle.\label{23} \ee Using the following decomposition of
$Y_{\mu\nu}$ and $K_{\mu\nu}$ in the trace and traceless parts \ba
Y_{\mu\nu}=\frac{Y}{5}q_{\mu\nu}+\varpi_{\mu\nu},\label{24}\\
K_{\mu\nu}=\frac{K}{5}q_{\mu\nu}+\zeta_{\mu\nu},\label{25}\ea the
projected equation on the brane reads \be
^{(5)}\!G_{\mu\nu}=-\bar{\Lambda}_{5}q_{\mu\nu}+8\pi
\Omega\tau_{\mu\nu}+8\Big(\frac{\pi}{\phi}\Big)^{2}\Sigma_{\mu\nu}+
\langle \varpi_{\mu\nu} \rangle+\frac{3}{5}\langle K \rangle
\langle \zeta_{\mu\nu} \rangle-\langle \zeta_{\mu}^{\;\;\alpha}
\rangle \langle \zeta_{\nu\alpha} \rangle,\label{26}\ee where
$\bar{\Lambda}_{5}$ is given by \be
\Lambda_{5}=\frac{3}{10}\langle Y \rangle+\frac{6}{25}\langle K
\rangle^{2}-\frac{1}{2}\langle \zeta^{\alpha\beta} \rangle \langle
\zeta_{\alpha\beta} \rangle+
\frac{3}{8}\Big(\frac{8\pi}{\phi}\Big)^{2}\frac{\lambda(w-1)}{(3+2w)^{2}}(\tau+\lambda(w-1)),\label{27}\ee
and
\begin{eqnarray}
\langle Y \rangle&=&\left.-2\Bigg(\Bigg\langle
\frac{8\pi}{\phi}T_{M\mu\nu}n^{\mu}n^{\nu} \Bigg\rangle
+\Bigg\langle\frac{w}{\phi^{2}}\Big(\phi,_{\mu}\phi,_{\nu}n^{\mu}n^{\nu}-\frac{1}{2}\phi,_{\alpha}\phi,
^{\alpha}\Big) \Bigg\rangle\right.\nonumber\\&+&\left.
\Bigg\langle\frac{1}{\phi}\Big(\phi,_{\mu
;\nu}n^{\mu}n^{\nu}-\frac{8\pi}{3+2w}T_{M}\Big)\Bigg\rangle\Bigg),\right.
\label{28}
\end{eqnarray} with $\phi,_{\mu}\equiv \nabla_{\mu}\phi$. Note the
appearance of $\langle \zeta_{\mu\nu} \rangle$-like terms in
(\ref{26}). According to the decomposition (\ref{25}) those terms
arise only due to the absence of the $\mathbb{Z}_{2}$ symmetry and
encodes information about the shear of the curled on-brane
dimension. Therefore, it seems that the application of the
Gauss-Codazzi formalism into non-$\mathbb{Z}_{2}$ symmetric
brane-worlds describes hybrid compactification scenarios in a more
natural way. Of course, by orbifolding the extra transverse
dimension one reobtains the previous Section results.

It is necessary to go one step further in order to determine the
mean value of the extrinsic curvature appearing in (\ref{26}). To
do so, let us define a convenient new brane matter stress-tensor
by $\hat{\tau}_{\mu\nu}
=\tau_{\mu\nu}+\frac{(3(1-w)\lambda-(w+3)\tau)}{4(3+2w)}q_{\mu\nu}$,
in terms of which the equation (\ref{19}) reads \be
0=[Y_{\mu\nu}]+\langle K \rangle [K_{\mu\nu}]+\frac{8\pi}{\phi}
\langle K_{[\mu}^{\;\;\alpha} \rangle \hat{\tau}_{\nu]\alpha}.
\label{29}\ee Now, after expressing $[K_{\mu\nu}]$ and $[K]$ in
terms of that new stress-tensor $\hat{\tau}_{\mu\nu}$ and inserting
it in equation (\ref{29}) we find \be \frac{8\pi}{\phi}\langle K
\rangle=\frac{3(\hat{\tau}^{-1})^{\mu\nu}
[Y_{\mu\nu}]}{9-(\hat{\tau}^{-1})^{\mu}_{\mu}\hat{\tau}^{\nu}_{\nu}}\label{30},\ee
and again from (\ref{29}) we arrive at \be -\frac{8\pi}{\phi}
\langle K_{[\mu}^{\;\;\alpha}\rangle \hat{\tau}_{\nu]\alpha}=
 [Y_{\mu\nu}]+\frac{3(\hat{\tau}^{-1})^{\alpha\beta}[Y_{\alpha\beta}]}{9-(\hat{\tau}^{-1})^{\sigma}_{\sigma}\hat{\tau}^{\gamma}_{\gamma}}
 (-\hat{\tau}_{\mu\nu}+\frac{\hat{\tau}}{3}q_{\mu\nu}),\label{31} \ee
or, in a more compact way, \be \frac{8\pi}{\phi} \langle
K_{[\mu}^{\;\;\alpha}\rangle \hat{\tau}_{\nu]\alpha}\equiv
-[\hat{Y}_{\mu\nu}].\label{nova}\ee The complete decoupling of
$\langle K_{\mu\nu}\rangle$ can be obtained from the vielbein
decomposition. Therefore, let us introduce a complete basis
$h_{\mu}^{(i)}$ $(i=0,1,...,4)$ of orthonormal vectors constructed
by the contraction of an orthonormal matrix set which represents a
local Lorentz transformation and turns $\hat{\tau}_{\mu\nu}$ (and
consequently $\tau_{\mu\nu}$) diagonal. The orthonormality
conditions are given by
\begin{eqnarray}
\left.h^{\mu}_{\;\;(i)}h_{\mu (j)}=\eta_{(i)(j)},\right.\nonumber \\
\sum_{i,j=0}^{4}\eta_{(i)(j)}h_{\mu}^{\;\;(i)}h_{\nu}^{\;\;(j)}=\sum_{j=0}^{4}h_{\mu}^{\;\;(j)}h_{\nu
(j)}=q_{\mu\nu},\label{31}
\end{eqnarray}
where $\eta_{(i)(j)}$ is the Minkowski metric\footnote{Note that
we are not assuming Einstein's summation convention over the
tangent indices.}. Expressing $\hat{\tau}_{\mu\nu}$ in terms of
the vielbein,
$\hat{\tau}_{\mu\nu}=\sum_{i}\hat{\tau}_{(i)}h_{\mu}^{\;\;(i)}h_{\nu
(i)}$, we have from (\ref{nova}) \be \frac{8\pi}{\phi}\langle
K_{\mu\nu}\rangle=-\sum_{i,j}
\frac{h_{\mu}^{\;\;(i)}h_{\nu}^{\;\;(j)}}{\hat{\tau}_{(i)}+\hat{\tau}_{(j)}}[\hat{Y}_{(i)(j)}],\label{32}
\ee after a contraction with $h^{\mu}_{\;\;(i)}h^{\nu}_{\;\;(j)}$.
In the equation (\ref{32}), $[\hat{Y}_{(i)(j)}]\equiv
h^{\mu}_{\;\;(i)}h^{\nu}_{\;\;(j)} [\hat{Y}_{\mu\nu}]$. So, since
the diagonal term of (\ref{32}) is given by
$\sum_{i=j}\frac{h_{\mu}^{\;\;(i)}h_{\nu}^{\;\;(j)}}{\hat{\tau}_{(i)}+\hat{\tau}_{(j)}}[\hat{Y}_{(i)(j)}]
=\frac{1}{2}(\hat{\tau}^{-1})_{\mu}^{\;\;\alpha}[\hat{Y}_{\alpha\nu}]$,
the generalized matching condition to the mean value of the extrinsic
curvature reads
\begin{eqnarray}
\frac{8\pi}{\phi}\langle K_{\mu\nu} \rangle
&=&\left.\frac{1}{2}(\hat{\tau}^{-1})_{\mu}^{\;\;\alpha}[Y_{\alpha\nu}]
+\frac{3(\hat{\tau}^{-1})^{\beta\gamma}[Y_{\beta\gamma}]}{2(9-(\hat{\tau}^{-1})^{\sigma}_{\sigma}\hat{\tau}^{\rho}_{\rho})}
\Big(q_{\mu\nu}-\frac{\hat{\tau}^{\rho}_{\rho}(\hat{\tau}^{-1})_{\mu\nu}}{3}\Big)\right.\nonumber\\
&-&\left.\sum_{i\neq
j}\frac{h_{\mu}^{\;\;(i)}h_{\nu}^{\;\;(j)}}{\hat{\tau}_{(i)}+\hat{\tau}_{(j)}}[\varpi_{(i)(j)}].\right.\label{33}
\end{eqnarray}

From equation (\ref{33}) one can find the expression for $\langle
K \rangle$, while from the decomposition (\ref{25}) one finds
$\langle \zeta_{\mu\nu} \rangle$. After all, the projected
Einstein-Brans-Dicke equation in the orthonormal frame has the
following diagonal terms \be
^{(5)}\!G_{(i)(i)}=-\bar{\Lambda}_{5}+8\pi
\Omega\tau_{(i)}+\Sigma_{(i)}+\langle \varpi_{(i)(i)}
\rangle+\frac{3}{5}\langle K \rangle \langle \zeta_{(i)(i)}
\rangle+\sum_{k}\langle \zeta_{(i)}^{\;\;(k)} \rangle \langle
\zeta_{(i)(k)} \rangle,\label{34} \ee where $\Sigma_{(i)}=
\frac{1}{4}\Bigg(\frac{8\pi}{\phi}\Bigg)^{2}\Bigg(\frac{(w+3)}{2(3+2w)}\tau\tau_{(i)}-
\tau_{(i)}^{2}+\frac{1}{2}\Big(\sum_{j}\tau_{(j)}^{2}\Big)-\frac{(w^{2}+3w+3)}{2(3+2w)^{2}}\tau^{2}\Bigg)$.
Moreover, the absence of the $\mathbb{Z}_{2}$ symmetry allows the
existence of off-diagonal terms of the Einstein brane tensor, given
by \be ^{(5)}\!G_{(i)(j)}=\langle \varpi_{(i)(j)}
\rangle+\frac{3}{5}\langle K \rangle \langle \zeta_{(i)(j)}
\rangle+\sum_{k}\langle \zeta_{(i)}^{\;\;(k)}\rangle \langle
\zeta_{(j)(k)} \rangle.\label{35} \ee the equations (\ref{34}) and
(\ref{35}) are the result of the generalization of the
Gauss-Codazzi formalism to brane-worlds without the orbifold
symmetry. A general characteristic of this procedure is the
appearance of terms proportional to the shear of the extrinsic
curvature, as well as the existence of off-diagonal elements in
the Einstein projected tensor. These last two properties enables
one to say that this formalism can extract more physical
information when applied to hybrid compactification models.

\section{Outlooks}

This work can be positioned in the middle road between the pure
formalism and the application. A more formal approach must take
into account the advices arising in the study of consistence
conditions for Brans-Dicke brane-worlds (BDBW) \cite{ULT}.
Nevertheless, the use of Gauss-Codazzi formalism is necessary in
order to construct a bridge between formalism and phenomenology.

Apart of the rather technical approach focused in this work, the
result encoded in the equations (\ref{9}), (\ref{34}) and
(\ref{35}) is exhaustive: the presence of the dilatonic field, by
all means, brings new signatures if applied to cosmological
systems. In this vein, several possibilities are open to further
investigation in the scope of BDBW. A systematic study of the
problems analyzed, for instance, in \cite{GALERA} in the context
of BDBW models can certainly provide new insights about high
energy physics, as well as about the physics of extra dimensions
itself. Moreover, we hope that in the study of specific
cosmological problems, as the galactic rotation curves for
example, the ubiquitous presence of the Brans-Dicke parameter
restricts the wide range of possible adjustments, coming from the
projected gravitational field equations, and points out to a more
phenomenologically viable scenario. Of course, such restriction is
possible only if one is willing to accept that the current lower
bound of the Brans-Dicke parameter, coming from Solar System
experiments \cite{EXP}, is also valid in the brane-world
framework.

To finalize we should remark that, since the Brans-Dicke theory
can mimic gravity recovered from string theory at low energy, at
least in some regimes, the study of the cosmological aspects of
BDBW can also bring some information about problems in string
cosmology.

\section*{Acknowledgments}

The authors are very grateful for the invitation to make a
contribution with this work. J. M. Hoff da Silva thanks to
CAPES-Brazil for financial support. M. C. B. Abdalla and M. E. X.
Guimar\~aes acknowledge CNPq for support.

\end{document}